# Sliding ferroelectric memories and synapses

Xiuzhen Li[1,2,8], Biao Qin[3,8], Yaxian Wang[1,2,8], Yue Xi[4], Zhiheng Huang[1,2], Mengze Zhao[3], Yalin Peng[1,2], Zitao Chen[1], Zitian Pan[1,2], Jundong Zhu[1,2], Chenyang Cui[1,2], Rong Yang[5], Wei Yang[1,2], Sheng Meng[1,2], Dongxia Shi[1,2], Xuedong Bai[1,2], Can Liu[6], Na Li[1,7], Jianshi Tang[4*], Kaihui Liu[3,7*], Luojun Du[1,2*], Guangyu Zhang[1,2,7*]

[1]*Beijing National Laboratory for Condensed Matter Physics and Institute of Physics, Chinese Academy of Sciences, Beijing, 100190, China.*
[2]*School of Physical Sciences, University of Chinese Academy of Sciences, Beijing, 100190, China.*
[3]*State Key Laboratory for Mesoscopic Physics, Frontiers Science Center for Nano-optoelectronics, School of Physics, Peking University, Beijing, 100871, China*
[4]*School of Integrated Circuits, Beijing Advanced Innovation Center for Integrated Circuits, Tsinghua University, Beijing, 100084, China*
[5]*College of Semiconductors (College of Integrated Circuits), Hunan University, Changsha, 410082, China*
[6]*Key Laboratory of Quantum State Construction and Manipulation (Ministry of Education), Department of Physics, Renmin University of China, Beijing, 100872, China*
[7]*Songshan Lake Materials Laboratory, Dongguan, Guangdong, 523808, China.*
[8]*These authors contributed equally: Xiuzhen Li, Biao Qin, Yaxian Wang.*
*Corresponding authors: luojun.du@iphy.ac.cn; jtang@tsinghua.edu.cn; khliu@pku.edu.cn; gyzhang@iphy.ac.cn

**Ferroelectric materials with switchable electric polarization hold great promise for a plethora of emergent applications, such as post-Moore's law nanoelectronics, beyond-Boltzmann transistors, non-volatile memories, and above-bandgap photovoltaic devices. Recent advances have uncovered an exotic sliding ferroelectric mechanism, which endows to design atomically thin ferroelectrics from non-ferroelectric parent monolayers. Although notable progress has been witnessed in understanding its fundamental properties, functional devices based on sliding ferroelectrics, the key touchstone toward applications, remain elusive. Here, we demonstrate the rewritable, non-volatile memory devices at room-temperature utilizing a two-dimensional (2D) sliding ferroelectric semiconductor of rhombohedral-stacked bilayer molybdenum disulfide. The 2D sliding ferroelectric memories (SFeMs) show superior performances with a large memory window of >8V, a high conductance ratio of above $10^6$, a long retention time of >10 years, and a programming endurance greater than $10^4$ cycles. Remarkably, flexible SFeMs are achieved with state-of-the-art performances competitive to their rigid counterparts and maintain their performances post bending over $10^3$ cycles. Furthermore, synapse-specific Hebbian forms of plasticity and image recognition with a high accuracy of 97.81% are demonstrated based on flexible SFeMs. Our work demonstrates the sliding ferroelectric memories and synaptic plasticity on both rigid and flexible substrates, highlighting the great potential of sliding ferroelectrics for emerging technological applications in brain-inspired in-memory computing, edge intelligence and energy-efficient wearable electronics.**

## Main text

A ferroelectric is a polar material that features spontaneous charge polarization with two or more preferred stable polarization states, determined by its lattice symmetry breaking[1,2]. The ferroelectric polarization, which can be programmed (switched) via applying an external electrical field, is appealing for a myriad of technological advances and novel functionalities, including but not limited to ultra-dense random-access memories, in-memory computing to break today's von Neumann bottleneck, negative capacitance transistors to beat the 'Boltzmann Tyranny', photovoltaic devices beyond the Shockley-Queisser limit, and artificial neurons for spiking neural networks[1,3-10].

Triggered by the growing technological demands of continued device downscaling and emergent data-centric applications, ferroelectricity down to atomically thin limit has become highly sought-after, but is extremely challenging to obtain in conventional materials with three-dimensional (3D) lattices[5,6]. Two-dimensional (2D) layered materials knitted together by weak van der Waals (vdW) interactions provide an unprecedented platform for atomically thin ferroelectrics. In just a few years, we have had at our disposal dozens of 2D vdW ferroelectrics, such as $CuInP_2S_6$, $In_2Se_3$, $WTe_2$, group-IV monochalcogenide monolayers and graphene-boron nitride heterostructure[5,6,11-14]. Of particular interest, recent advances in 2D vdW ferroelectrics have uncovered a fundamentally new driving mechanism behind the spontaneous polarization schemes, i.e., sliding ferroelectrics[15-28]. Such exotic sliding ferroelectric mechanism enables the on-demand design of atomically thin ferroelectrics from non-ferroelectric parent monolayers, which can not only significantly broaden the range of available 2D ferroelectric materials (such as the scarce ferroelectric semiconductors), but also potentially outline a bright vision for a broad portfolio of emerging technological innovations. Although notable progress in 2D sliding ferroelectrics has been witnessed over the past few years, most studies are focused on identifying the presence of sliding ferroelectricity. Achieving functional sliding ferroelectric devices with competitive performances, which lays a strong foundation for 2D sliding ferroelectrics to fit into practical applications, is promising on the horizon yet remains elusive.

Here, we report the sliding ferroelectric field-effect transistors (FETs) at room-temperature in which a 2D sliding ferroelectric semiconductor of rhombohedral-stacked (R-stacked) bilayer $MoS_2$ grown by a stacking-controlled technique is employed as the channel material. The fabricated sliding ferroelectric transistors at thicknesses of only two atomic layers exhibit rewritable, non-volatile memory behaviours with excellent performances, including a large memory window of >8V, a high on/off conductance ratio of ~$10^7$, a long retention time of >10 years, and a programming endurance of over $10^4$ cycles. Importantly, flexible sliding ferroelectric memories (SFeM) are achieved with state-of-the-art performances competitive to their rigid counterparts and can be bent more than $10^3$ times without degrading device performances. Further, flexible electronic synapses which enable the emulation of specific Hebbian forms of plasticity, such as long-term potentiation and depression, are demonstrated with sliding ferroelectric bilayer $MoS_2$ transistors. Finally, we construct a five-layer convolutional neural network (CNN) based on flexible sliding ferroelectric synapses and achieve an accuracy as high as 97.81% in the recognition of MNIST (Modified National Institute of Standards and Technology) handwritten-digit images.

**Sliding ferroelectrics of 3R-stacked bilayer $MoS_2$**

Monolayer $MoS_2$, one of the most promising 2D semiconductors, holds great prospects for next-generation integrated circuits and novel flexible electronics because of its high intrinsic carrier mobility, excellent gate controllability, high on/off current ratio, ultra-low standby power, and remarkable flexibility[29-33]. Although inversion symmetry is intrinsically broken, monolayer $MoS_2$ (point group $D_{3h}$) is non-ferroelectric because of the out-of-plane mirror symmetry (upper panel of Fig. 1a). Remarkably, by stacking two $MoS_2$ monolayers in a parallel manner but with a lateral shift of one-third of the unit cell along the armchair direction, a R-stacked bilayer structure (polar point group $C_{3v}$) emerges, which breaks both the inversion and out-of-plane mirror symmetries (lower panel of Fig. 1a). This enables the generation of a spontaneous out-of-plane charge polarization from a general symmetry-breaking standpoint[2]. In addition, the R-stacked bilayer $MoS_2$ features two energy-degenerate stacking configurations, i.e., XM (MX) stacking order corresponding to the vertical alignment of chalcogen atoms in the top layer with molybdenum atoms in the bottom layer (and vice versa). Because such two stacking configurations can be viewed as mirror images of each other, their corresponding electric polarizations are opposite. Consequently, swapping the stacking configurations from one to another by an in-plane shear motion could result in the switching of spontaneous out-of-plane polarization and therefore exotic sliding ferroelectricity (lower panel of Fig. 1a).

Previously, the presence of sliding ferroelectric polarization in R-stacked bilayer $MoS_2$ has been demonstrated in micron-scale samples obtained by direct mechanical exfoliation or assembled by the tear and stamp technique[19,20,22,23,34,35]. Here we employ inch-scale R-stacked bilayer $MoS_2$ grown by a stacking-controlled technique[36]. Inset of Fig. 1b displays the optical microscopy image of R-stacked bilayer $MoS_2$ transferred onto a silicon substrate with 20 nm $Al_2O_3$ dielectric layer for device fabrication which we shall come to shortly. Apparently, the transferred sample is quite uniform and clean, devoid of crackles and bubbles. The height image obtained from a tapping mode atomic force microscopy (AFM) is presented in Fig. 1b. The height line scan profile indicates that the thickness of $MoS_2$ is ~1.27 nm, providing a fingerprint of the bilayer nature. This is further corroborated through Raman spectra with a frequency difference between $E_{2g}$ and $A_{1g}$ phonon modes of ~23.6 cm$^{-1}$ (top panel, Fig. 1c)[37].

To verify that our bilayer $MoS_2$ belongs to R-stacking phase with sliding ferroelectricity rather than the trivial centrosymmetric 2H-stacking structure, we perform optical second harmonic generation (SHG) measurements (bottom panel of Fig. 1c). The SHG intensity of our bilayer sample is around 4 times that of monolayer. This largely evidences the R-stacking configuration given that the SHG efficiency from R-stacked $MoS_2$ shows a quadratic dependence on the layer number as a result of atomically phase-matched conditions[38,39]. By contrast, the SHG response of 2H-stacked bilayer $MoS_2$ would vanish due to the existence of inversion symmetry. High-angle annular dark field scanning transmission electron microscopy (HAADF-STEM) is employed to further examine the stacking order. Atomic-resolution HAADF-STEM images (Fig. 1d and Supplementary Fig. 1) show that atoms are located not only at the corners of honeycomb lattice, but also at its center. This matches well with the R-stacked bilayer structure, but is incompatible with the 2H phase where atoms are located only at the corners of hexagonal lattice (Supplementary Fig. 2). We highlight that our samples are pure R-stacked bilayers (Supplementary Fig. 3), in stark contrast to a recent study in which R-stacked bilayer $MoS_2$ domains are only a small percentage and embedded in a highly dislocated and unstable non-ferroelectric matrix[40].

Having identified the R-stacked bilayer structure, we then perform piezoresponse force microscopy (PFM) measurements in the DART-SS-PFM mode to investigate the sliding ferroelectric behaviour and spontaneous charge polarization switching. Figure 1e displays the off-field PFM phase (top panel) and amplitude (bottom panel) against the constant-voltage bias for an as-grown R-stacked bilayer $MoS_2$ directly on the metal substrate. The PFM amplitude exhibits a characteristic butterfly loop, while the corresponding PFM phase shows a hysteresis loop with a phase switching of approximately 180° at the minima of the amplitude response, evidencing the electrically switchable polarization states and ferroelectric nature. The existence of ferroelectricity in pure R-stacked bilayer $MoS_2$ is highly consistent with the recent optical spectroscopy imaging results[34] and the observed ferroelectricity in untwisted, commensurate $MoS_2$/$WS_2$ heterobilayers[21]. To further verify the existence of ferroelectricity in R-stacked bilayer $MoS_2$, we perform *ab initio* density functional theory (DFT) calculations to extract the interfacial differential charge densities (DCD, a quantity measuring the charge variation at the interface) of the two energy-degenerate stacking configurations[21]. Left panel of Fig. 1f (1g) presents the calculated interfacial DCD for XM (MX) stacking order. The separation of the electron accumulation (rose red) and the electron depletion regions (green) illustrates the explicit charge redistribution between the top and bottom layers. Their line profiles (right panels of Figs. 1f and 1g) unequivocally show the spontaneous out-of-plane electric polarization at the interfaces and its dipole direction can be programmed by switching the stacking configurations from one to another through a 1.23 Å lateral sliding along the armchair direction.

**Sliding ferroelectric memory**

R-stacked bilayer $MoS_2$ belongs to an exotic sliding ferroelectric semiconductor integrating the superior transistor performances of non-ferroelectric monolayer $MoS_2$ and robust ferroelectric properties designed from sliding mechanism, and thus is expected to hold great promise for emerging technological applications such as non-volatile memories, neuromorphic computing, and brain-inspired intelligent devices[7,41,42]. To demonstrate such potentials, we first fabricate sliding ferroelectric FET devices utilizing R-stacked bilayer $MoS_2$ as the channel material. Benefitting from the large-scale R-stacked bilayer $MoS_2$ afforded by the direct growth method, the batch fabrication of sliding ferroelectric FET arrays over several millimeters is realized (left panel of Fig. 2a; Methods provides the fabrication details). Right panel of Fig. 2a illustrates the device geometry with Ti-Au-Ti as the local back-gate electrode, 20-nm $Al_2O_3$ as the dielectric layer, and R-stacked bilayer $MoS_2$ as the ferroelectric semiconductor channel. Cross-sectional STEM image of a typical device can be found in Supplementary Fig. 4. Semimetallic Sb is chosen as the source/drain electrodes to suppress Fermi-level pinning and reduce contact resistance[43]. Unless otherwise specified, all electrical measurements in the main text are performed at room temperature with an Agilent semiconductor parameter analyzer under a base pressure of $10^{-6}$ Torr.

Figure 2b plots the $I_{DS}$-$V_{DS}$ output characteristic curves of a representative R-stacked bilayer $MoS_2$ FET device under different back gate voltages $V_{GS}$ (sweeping from 0 V to 10 V with a step of 1 V). The overall linear output characteristics suggest the low-resistance Ohmic contact behaviour. Figure 2c presents the double-sweep $I_{DS}$-$V_{GS}$ transfer characteristics of a typical R-stacked bilayer $MoS_2$ FET device by scanning the back gate $V_{GS}$ from −10 V to +10 V and then back to −10 V at a fixed drain voltage $V_{DS}$ = 1 V. The R-stacked bilayer $MoS_2$ FET shows a typical n-type semiconducting characteristic with a high on/off current ratio of ~5×10$^7$. Remarkably, a prominent hysteretic loop is observed. To confirm that the observed $I_{DS}$-$V_{GS}$ hysteretic loop is from the switching of sliding ferroelectric polarization rather than from extrinsic origins such as trapped charges at the interface, we fabricate non-ferroelectric monolayer $MoS_2$ FET devices adopting exactly the same manufacturing process. Inset of Fig. 2c and Supplementary Fig. 5 plot the bidirectional scanning $I_{DS}$-$V_{GS}$ transfer curves of monolayer $MoS_2$ devices. The $I_{DS}$-$V_{GS}$ hysteresis is negligible for all non-ferroelectric monolayer $MoS_2$ FETs when the gate voltage is scanned in a closed loop between −10 and +10 V (more than 20 devices are measured). The strong contrast between the results of ferroelectric bilayer $MoS_2$ and non-ferroelectric monolayers provides the smoking gun evidence that the observed hysteresis windows in R-stacked bilayer $MoS_2$ originate from the ferroelectric polarization switching.

Figure 2d displays the bidirectional scanning $I_{DS}$-$V_{GS}$ transfer curves of a R-stacked bilayer $MoS_2$ FET device under different $V_{GS}$ sweeping ranges, where the $V_{DS}$ is fixed at 1 V. The extracted hysteresis memory window (ΔV) at $I_{DS}$ = 1 nA against the applied maximum gate voltage ($V_{max}$) is presented in the top panel of Fig. 2e. Obviously, no noticeable hysteresis window is observed when the $V_{max}$ is less than 4 V. This is highly consistent with the ferroelectric feature that the switching of polarization state typically requires a critical electric field. When the $V_{max}$ is ≥ 4 V, the hysteresis window becomes larger as the sweeping range $V_{GS}$ increases. This indicates that the ferroelectric polarization can be precisely programmed by the applied electric fields, facilitating multiple independent drain-to-source conductance states for electronic synapses as will be shown in latter discussions. A large hysteresis memory window of ~4.6 V is obtained under a $V_{GS}$ sweep range of ±10 V. It is noteworthy that the memory window can be further enlarged by increasing the thickness or lowering the dielectric constant of dielectric layer, but this will inevitably lead to an increase in power consumption. Therefore, a trade-off between enlarged memory window and minimized power consumption is required based on the practical applications.

As a result of the $I_{DS}$-$V_{GS}$ hysteresis, the R-stacked bilayer $MoS_2$ FET device exhibits two distinct resistance states at zero gate voltage ($V_{GS} = 0$), i.e., low resistance state (LRS) and high resistance state (HRS). The conductance ratio between the two states versus $V_{max}$ is depicted in the bottom panel of Fig. 2e. Apparently, when the applied $V_{max}$ is larger than 4 V, the conductance ratio increases sharply with $V_{max}$ and can reach up to ~8×10$^6$ under a $V_{GS}$ sweeping range of ±10 V. The retention characteristics of R-stacked bilayer $MoS_2$ FET devices reveal that both the LRS and HRS do not show obvious degradation up to $10^3$ s (Supplementary Fig. 6a), indicating stable non-volatile memory behaviour. More remarkably, the extrapolation of data retention suggests that the LRS and HRS could be maintained for more than 10 years with a conductance ratio exceeds $10^6$. Further, the programming endurance characteristics demonstrate that the R-stacked bilayer $MoS_2$ FET devices can survive over ~10$^4$ cycles without degradation (Supplementary Fig. 6b). The extrapolation of endurance indicates that negligible LRS and HRS degradation and a conductance ratio of above $10^6$ can potentially be remained after $10^{13}$ switching cycles, which is adequate for memory applications and can even meet the requirement for edge ($10^9$) and cloud training ($10^{12}$)[4,44]. Our SFeM integrates excellent retention time and superb programming endurance, promising the prospect of data-centric applications such as in-memory computing[4,44].

To evaluate the device uniformity, a 5×5 R-stacked bilayer $MoS_2$ FET array is tested. By programming the sliding ferroelectric polarization states with a gate voltage of +10 V (i.e., HRS) or −10 V (i.e., LRS), non-volatile memory maps with the characters 'S', 'F', 'e' and 'M' are achieved successfully in succession by recording the conductivity at $V_{DS} = 1$ V and $V_{GS} = 0$ (Fig. 2f). The re-programmable memory maps and strong conductance contrast between HRS (blue) and LRS (red) evidence superior device homogeneity.

**Flexible sliding ferroelectric memory**
Thanks to the 2D nature of only two atomic layers, R-stacked bilayer $MoS_2$ would not only harbour robust ferroelectric properties, but also show remarkable flexibility[29,30]. Consequently, R-stacked bilayer $MoS_2$ may offer unprecedented opportunities to implement flexible ferroelectric memory and thus promises the prospect of emerging portable, wearable and implantable electronics[45]. We now pursue the fabrication of flexible non-volatile memory devices based on the sliding ferroelectric semiconductor of R-stacked bilayer $MoS_2$ in a back-gate geometry, where polyethylene terephthalate (PET) and 30-nm $Al_2O_3$ are employed as the substrate and dielectric layer, respectively (Fig. 3a). Figure 3b presents the dual-sweep $I_{DS}$-$V_{GS}$ transfer characteristics of a flexible R-stacked bilayer $MoS_2$ FET device under three different test conditions, i.e., the pristine flat state before bending (black), the bending state corresponding to a tensile strain of ~1.3% (red) and the post-bend flat state after 1000 consecutive cycles of bending test (blue). Apparently, the FET performances under these three test conditions such as on-off current ratio are essentially the same. This proves that R-stacked bilayer $MoS_2$ is remarkably flexible with an excellent tolerance to thousands of bending tests, facilitating the development of high-performance flexible devices.

Remarkably, a large hysteresis memory window ($\Delta V > 8$ V at $I_{DS} = 1$ nA) with the non-volatile LRS and HRS at $V_{GS} = 0$ differing by more than six orders of magnitude, is observed for all the three test conditions (Fig. 3b). This bears striking resemblance to the results measured on rigid silicon substrate (Fig. 2). It is noteworthy that the hysteresis memory window of flexible SFeM is larger than that of SFeM on rigid substrate under a same $V_{GS}$ sweeping range of ±10 V (~4.6 V, top panel of Fig. 2e). This can be understood as that the thickness of $Al_2O_3$ dielectric layer is thicker for flexible SFeM (i.e., 30 and 20 nm for SFeM on flexible and rigid substrate, respectively). Figure 3c depicts the retention characteristics of a flexible SFeM, which is pre-programmed to the HRS (LRS) through a +10 V (−10 V) gate voltage pulse with width of 100 ms, and then read out via a 1 V source-drain bias under a floating gate configuration. For all the three configurations measured, no appreciable degradation is observed up to ~$10^4$ s for both the non-volatile LRS and HRS, offering the affirmative evidence of a long data retention time. Significantly, even after being subjected to $10^3$ cycles of bending, we could still extrapolate the ten-year retention time for the flexible SFeM with a conductance ratio of ~$10^6$ between the LRS and HRS (inset of Fig. 3c).

To evaluate the endurance characteristics of flexible SFeM, we alternately apply programming gate voltage pulses of ±10 V (width, 100 ms; period, 500 ms), and monitor the HRS/LRS conductance at intervals of several cycles with a 1 V source-drain bias at $V_{GS}$ = 0 (Fig. 3d). Remarkably, the flexible SFeM does not show noticeable degradation after ~$10^4$ cycles of ferroelectric switching (Fig. 3d and Supplementary Fig. 7b). More importantly, the projected endurance indicates that the non-volatile LRS and HRS, differing by a conductance ratio of above $10^6$, can potentially maintain after $10^{13}$ programming cycles even after $10^3$ cycles of bending (Supplementary Fig. 7b). The endurance of flexible SFeM against the bending cycles is also evaluated (inset of Fig. 3d). The negligible HRS and LRS degradation up to $10^3$ cycles of bending reveals the superior immunity to bending. This reinforces the robustness of sliding ferroelectricity in R-stacked bilayer $MoS_2$ and will boost technological applications in a series of flexible scenario. We highlight that this is the first time, to the best of our knowledge, that flexible SFeM is constructed by employing ferroelectric semiconductor as the channel material.

To characterize the homogeneity, a 5×5 SFeM array is fabricated on flexible PET substrate. The conductances of the LRS and HRS of these 25 flexible SFeM devices after $10^3$ cycles of bending (the corresponding memory window and conductance ratio) are depicted in Fig. 3e (Fig. 3f), demonstrating good uniformity. Further, we benchmark the performances of our flexible SFeM. Overall, our flexible SFeM based on R-stacked bilayer $MoS_2$ shows competitive performances compared with their rigid counterparts, including a large memory window of >8V, a high conductance ratio of above $10^6$, a long retention time of >10 years, and a programming endurance greater than $10^4$ cycles. Figure 3g and Supplementary Table 1 compare the characteristics of our flexible SFeM (i.e., memory window, retention, endurance, conductance ratio, flexibility, and thickness) with those of previously reported ferroelectric FET devices with 2D ferroelectric materials as the channel materials or non-ferroelectric 2D semiconductor/ferroelectric materials as the channel materials/gate insulators, showing that our flexible SFeM at thicknesses of only two atomic layers is very promising[7,41,42,46-48].

**Flexible sliding ferroelectric synapses**
Typically, in human neural networks, the transmission of nerve impulse signals from a pre-synaptic to a post-synaptic neuron as neurotransmitters diffuse through the synaptic cleft can be recognized by an excitatory/inhibitory post-synaptic current (PSC) (Fig. 4a). The ability to modulate the PSC amplitude (i.e., synaptic weight), which is believed to endow the brain with the ability to learn and process information, is referred to as synaptic plasticity[49,50]. Effective emulation of synaptic-like behaviour in electronic devices, especially synaptic plasticity, would facilitate the development of energy-efficient brain-like neuromorphic computing to overcome the limitations of current von Neumann architectures in terms of energy consumption and throughput[4,49-53].

Given the precisely programmable ferroelectric polarization states, the SFeM based on R-stacked bilayer MoS$_2$ shows highly controllable drain-to-source conductance states with the magnitude and width of the gate voltage pulse (Supplementary Figs. 8 and 9), and thus can operate as an extraordinary artificial synapse to emulate the plasticity learning rules of biological synapses, where gate voltage pulses simulate pre-synaptic input and channel conductance is identified as the synaptic weight. More importantly, sliding ferroelectric synapses based on R-stacked bilayer MoS$_2$ can be achieved on not only rigid substrates but also on flexible substrates, as the excellent flexibility demonstrated above. It is noteworthy that flexible electronic synapses down to 2D limit (i.e., < 5 nm), which are in high demand for emerging wearable devices and edge intelligence devices, have not been implemented previously.

Figure 4b depicts the synaptic weight PSC of a SFeM on flexible PET substrate obtained by applying a gate voltage pulse sequence consisting of alternating 50 negative (−5 V, 15 ms) and 50 positive pulses (+5 V, 15 ms). Interestingly, via training with a consecutive negative (positive) gate pulse, the synaptic weight PSC increases (decreases) progressively, enabling the emulation of the long-term potentiation (long-term depression). Note that long-term potentiation and depression are two essential synapse-specific Hebbian forms of plasticity that underly the experience-based learning. To further demonstrate the neuromorphic computing of flexible SFeM, 32 independent drain-to-source conductance states are extracted (Fig. 4c). These conductance states are well separated and show good retention with no apparent degradation over time (Supplementary Fig. 10a), facilitating weight update and high accuracy. To this end, we construct a five-layer CNN based on flexible sliding ferroelectric synapses to execute the recognition of MNIST handwritten-digit images. Figure 4d illustrates the structure of the CNN containing two convolutional layers, two maximum-pooling layers, and a fully-connected layer. The detailed data flow in the CNN is described in Methods. After the CNN is properly trained by software which gives an accuracy baseline of 97.85%, the weights in convolutional kernel and fully-connected layer are quantized to 32 levels and transferred to flexible sliding ferroelectric synapses for subsequent edge inference task of MNIST handwritten-digit image recognition. Remarkably, the output maintains a high recognition accuracy of 97.81%, which is only a 0.04% loss compared to software baseline (Supplementary Fig. 10b). In addition, the high accuracy of handwritten-digit image recognition does not show obvious degradation over time (Supplementary Fig. 10b), indicating the superior retention and high fidelity of the edge inference results.

**Conclusion**
We demonstrate the room-temperature sliding ferroelectric non-volatile memories and synapses on both rigid and flexible substrates employing an as-grown inch-scale rhombohedral-stacked bilayer MoS$_2$ as the device channel material. The sliding ferroelectric memories exhibit excellent performances, including a large memory window of >8V, a high on/off conductance ratio of ~$10^7$, a long retention time of >10 years, and a programming endurance of over $10^4$ cycles. Because of the remarkable flexibility afforded by the 2D nature, flexible sliding ferroelectric memories with state-of-the-art performances competitive to their rigid counterparts are achieved and can maintain their performances post bending over $10^3$ cycles. Further, flexible sliding ferroelectric synapses which enable the emulation of specific Hebbian forms of long-term plasticity are demonstrated. Finally, we construct a five-layer CNN based on flexible sliding ferroelectric synapses and achieve an accuracy as high as 97.81% in the recognition of MNIST handwritten-digit images. Our work represents an important advancement in the field of 2D sliding ferroelectrics in terms of the realization of high-performance functional devices. Together with the unprecedented possibilities of on-demand design of atomically thin ferroelectrics (including ferroelectric semiconductors, dielectric insulators, and even metals) facilitated by sliding ferroelectric mechanism, we anticipate that sliding ferroelectrics will promise the prospect of emerging technological applications in brain-inspired in-memory computing, neuromorphic computing, edge intelligence and energy-efficient wearable electronics.

**Methods**

**Device fabrication and measurement.** 15 nm Ti (3 nm)/Au (10 nm)/Ti (2 nm) is patterned and deposited on silicon substrate as local back gate. Then, 20 nm $Al_2O_3$ is deposited by ALD as dielectric insulator at 200 °C using $H_2O$ and trimethyl aluminium as precursors. Afterwards, the as grown R-stacked bilayer $MoS_2$ is transferred from the growth substrate onto $Al_2O_3$ by using polydimethylsiloxane (PDMS). Channel is then patterned by lithography and reactive ion etching (RIE). Contact electrodes are defined by lithography and 10 nm antimony (Sb) and 30 nm gold (Au) are finally deposited. Flexible devices share similar fabrication process where rigid Si substrates are replaced by flexible PET and the ALD temperature is lower to 150 °C and thickness of dielectric layer is 30 nm. All the patterning process are done in an electron beam lithography system (Eline, Raith) with poly (methyl methacrylate) (PMMA) as photoresist and metal electrodes are deposited in a custom-built electron beam deposition system and followed by lift-off process. Electrical performances are measured in a custom-built probe station with semiconductor parameter analyser (B1500A, Agilent) under a base pressure of $10^{-6}$ Torr at room temperature.

**Optical and morphology characterization.** Optical and morphology characterizations are conducted after the sample is transferred onto a rigid silicon substrate with 20-nm $Al_2O_3$ as the dielectric layer. Optical properties of the sample are characterized by SHG measurement and Raman spectroscopy. SHG is performed on a WITec alpha 300R system with a laser excitation wavelength of 1064 nm. Raman spectra are acquired using a HORIBA spectrometer (LabRAM HR Evolution) in a confocal backscattering configuration through a grating of 1800 gr/mm. Light from 532 nm (2.33 eV) continuous laser with a power of 137 μW for Raman measurements is focused through a Nikon objective (N.A.=0.5, W.D.=10.6, F.N.=26.5) onto the sample with a spot diameter of ~ 2 μm. The spectrometer integration time is 30 s. Morphology characterization is carried out on an AFM (Cypher, Oxford instruments) under tapping mode, using the tip AC240TS-R3 (Oxford instruments).

**PFM measurement.** PFM measurements are conducted in-situ on the epitaxy metal substrate using DART-SS-PFM mode to ensure strong signal-to-noise ratio on a commercial scanning probe microscope system (Cypher, Oxford instruments). DART stands for "Dual Amplitude Resonance Tracking", the SS stands for "Switching Spectroscopy". Antimony doped and Pt/Ir coated conductive silicon tips (SCM-PIT-V2, BRUKER) are utilized. The PFM hysteresis and loops are measured conducting on-off field switching technique. The external fields between tip and Ni during "on" stage switch the ferroelectric polarization of the sample while the piezoresponses are measured at "off" state where the field is canceled to avoid electrostatic effects.

**STEM characterization.** High-angle annular dark-field scanning transmission electron microscopy (HAADF-STEM) and STEM energy-dispersive X-ray spectroscopy (HAADF-STEM-EDX) are conducted on JEM-ARM200CF system (JEOL) operating at 200kV. Samples for top view stacking characterization are transferred onto a STEM grid. Samples for cross-section characterization and EDX analysis are cut from devices on silicon substrate by focused-ion-beam (FIB) system (Helios 600I, FEI).

**CNN simulation.** A series of simulations are performed for the application of MNIST recognition. The simulator consists of a five-layer CNN. As depicted in Fig. 4d, the input is a 28 × 28 greyscale (8-bit) digit image. The first convolutional layer (C1) measuring 26 × 26 × 16 (weight × height × depth) is obtained after convolution with the 1 × 3 × 3 × 16 kernels (depth × weight × height × batch). The result is subsampled by a pooling layer (P1, 9 × 9 × 16) utilizing a 3 × 3 maximum-pooling operation with a sliding stride of 3. Subsequently, the second convolutional layer (C2, 7 × 7 × 32) is acquired with 32 stacked feature maps by a secondary convolution with the 16 × 3 × 3 × 32 kernels. This is followed by another round of subsampling, executed by another pooling layer (P2, 4 × 4 × 32) using a 2 × 2 maximum-pooling operation with a stride of 2. Ultimately, the expanded 512-element vector is applied to a fully-connected layer as an input, aiming to attain the computing result in accordance with the 10 probability outputs. The network training was completed by software with a 97.85% baseline accuracy. Throughout the course of these simulations, the measured data was normalized into [-1,1] according to its conductance range, and then replaced the 4-bit or 5-bit quantized weights in both kernel and FC layer. The accuracy over time is directly related to the change of device conductance value.

**DFT calculations.** We perform density functional theory (DFT) calculations using the Vienna ab-initio simulation package (VASP)[54,55] with the exchange-correlation potential described by the generalized gradient approximation (GGA) and the Perdew-Burke-Ernzerhof (PBE) functiona[56]. We employ the projector augmented-wave (PAW) method[57] and a plane-wave basis with kinetic energy cut-off of 520 eV. A Monkhorst-Pack $k$-mesh of 6×6×1 was adopted to sample the first Brillouin zone[58]. We used a vacuum layer of 21 Å to exclude the interactions from adjacent image layers, and applied a dipole correction along the z-direction to correct the artificial electric polarization introduced by the periodic boundary condition. We calculated the differential charge density using VASPKIT package[59] by the difference between the total charge density of the bilayer and the superposition of the charge densities of two separate single layers, and the charge transfer was obtained as a planer integration of the differential charge density.

**Acknowledgements**
We thank Dr. Qing Yang for valuable discussions, and thank Yuqing Feng, Qinghua Zhang, Huanfang Tian and Aizi Jin for help on data processing, fabrication, and characterization. We acknowledge supports from the National Key Research and Development Program of China (Grant Nos. 2021YFA1202900, 2021YFA1400502, 2020YFA0309600, 2022YFA1403504), National Science Foundation of China (NSFC, Grant Nos. 12274447, 11834017, 62204166, 12074412, 52025023), the Strategic Priority Research Program of CAS (Grant Nos. XDB0470101) and Guangdong Major Project of Basic and Applied Basic Research (2021B0301030002).

**Author contributions**
L.D. and G.Z. supervised the research project. X.L. and L.D. conceived the idea of constructing sliding ferroelectric memory on R-stacked bilayer $MoS_2$. X.L. carried out PFM measurements and device fabrication and measurement. B.Q. grown high-quality R-stacked $MoS_2$ samples and conducted SHG measurements under the supervision of C.L and K.L.. Y.W. carried out the DFT calculation. Y.X. conducted CNN simulation under the supervision of J.T.. Z.H. performed Raman spectra. Y.P. and C.C. contributed to device measurements. Z.P. and J.Z. prepared TEM samples. Z.C. and X.B. carried out the TEM measurements. M.Z. contributed to the schematics. L.D., X.L., G.Z. wrote the manuscript with the input from all authors.

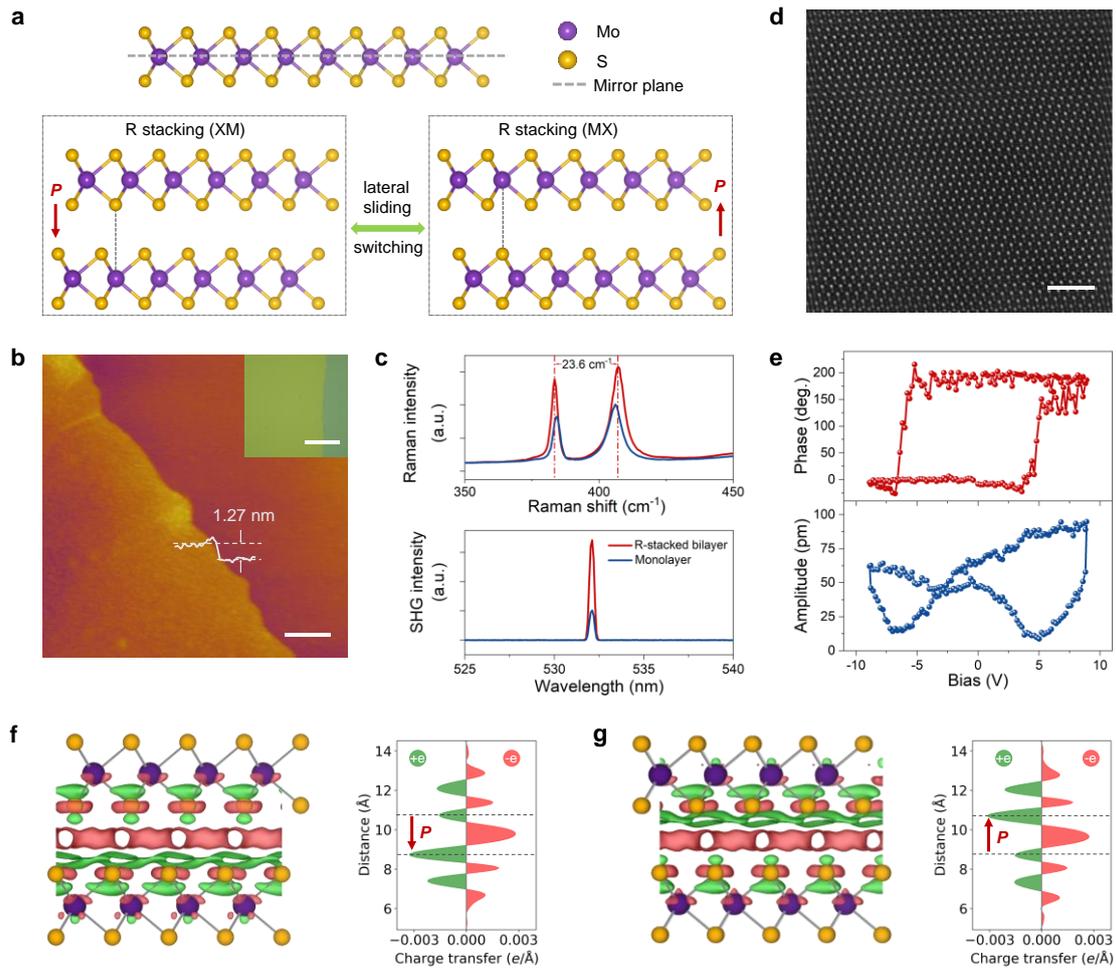

**Fig. 1 | Sliding ferroelectricity of 3R-stacked bilayer MoS$_2$. a**, Cross-section view along the armchair direction of monolayer MoS$_2$ (top) and R-stacked bilayer MoS$_2$ (bottom). R-stacked bilayer MoS$_2$ features two energy-degenerate stacking configurations, i.e., XM (MX) stacking order corresponding to the vertical alignment of sulfide atoms in the top layer with molybdenum atoms in the bottom layer (and vice versa). **b**, AFM height image of R-stacked bilayer MoS$_2$ showing thickness of around 1.27 nm; scale bar 500 nm. Inset: optical image of R-stacked bilayer MoS$_2$ after transferring onto a silicon substrate coated with 20 nm Al$_2$O$_3$; scale bar 200 μm. **c**, Raman (up) and SHG spectra (bottom) of monolayer (light blue) and R-stacked bilayer MoS$_2$ (red). **d**, Atomic resolution HAADF-STEM image of R-stacked bilayer MoS$_2$. Scale bar, 2 nm. **e**, Off-field PFM phase (top) and amplitude (bottom) of an as-grown R-stacked bilayer MoS$_2$ against the constant-voltage bias directly on the metal substrate, showing the typical ferroelectric hysteresis and butterfly loop, respectively. **f**, **g**, Interlayer charge density distributions (left) and their corresponding line profiles along *z* direction (right) for XM (**f**) and MX stacking orders (**g**), indicating opposite ferroelectric polarizations between them.

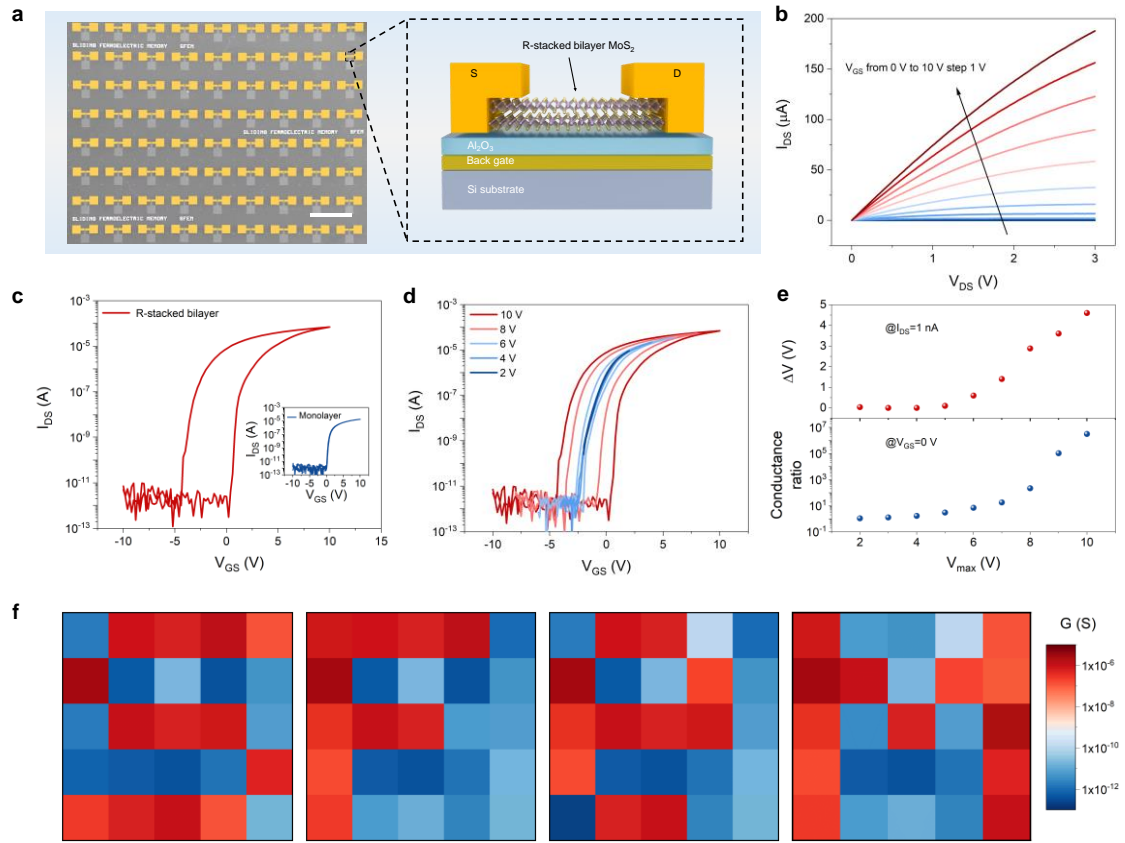

**Fig. 2 | Sliding ferroelectric memory on rigid substrate. a**, Left: false color scanning electron microscopy image of sliding ferroelectric FET arrays. Scale bar, 500 μm. Right: schematic of the device structure which consists of rigid silicon substrate, local metal gate, 20 nm $Al_2O_3$ dielectric insulator, R-stacked bilayer $MoS_2$ channel, and metal contacts. **b, c,** Output (**b**) and dual-sweep transfer characteristic curves (**c**) of a R-stacked bilayer $MoS_2$ device. Inset in (**c**) is the transfer curves of a monolayer $MoS_2$ device. **d**, Dual-sweep transfer curves of R-stacked bilayer $MoS_2$ devices with different scanning voltages. **e**, Hysteresis memory window at $I_{DS}$=1 nA (top) and conductance ratio at $V_{GS}$=0 V (bottom) as a function of the sweeping voltages. **f**, Colour maps of conductance for a 5×5 R-stacked bilayer $MoS_2$ FET arrays recorded at $V_{DS}$ = 1 V and $V_{GS}$ = 0. By programming the sliding ferroelectric polarization states with a gate voltage of +10 V (i.e., HRS) or −10 V (i.e., LRS), non-volatile memory maps with the characters 'S', 'F', 'e' and 'M' are achieved in succession.

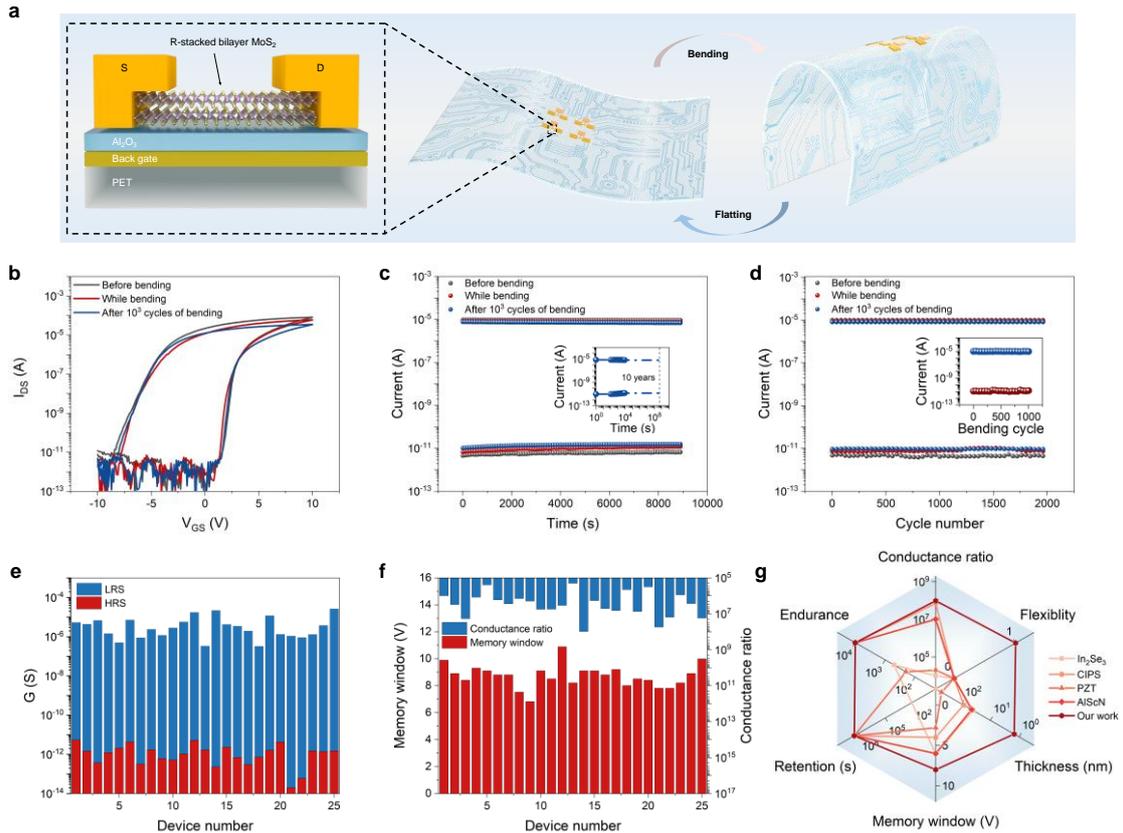

**Fig. 3 | Flexible sliding ferroelectric memory. a**, Box: Schematic of flexible SFeM device on PET substrate. Illustration of flexible SFeM under flat (middle) and bending (right) states. **b**, Transfer characteristic curves of flexible SFeM under different test conditions, i.e., the pristine flat state before bending (red), the bending state corresponding to a tensile strain of ~1.3% (red) and the flat situation after 1000 consecutive cycles of bending test (blue). **c**, Retention characteristics of a flexible SFeM, which is pre-programmed to the HRS (LRS) through a +10 V (−10 V) gate voltage pulse with width of 100 ms, and then read out via a 1 V source-drain bias at $V_{GS} = 0$. Inset: the extrapolation of the data retention indicates that the LRS and HRS could be maintained for more than 10 years with a conductance ratio exceeds $10^6$ even after being subjected to $10^3$ cycles of bending. **d**, Endurance characteristics of a flexible SFeM. The programming gate voltage pulses are alternating +10 V and −10 V in amplitude and 100 ms in width. Inset: the endurance of a flexible SFeM against the bending cycle. **e, f,** Statistical data of conductance of LRS/HRS (**e**) and conductance ratio/memory window (**f**) from 25 devices after 1000 cycles of bending. **g**, Comparation of our flexible SFeM with previously reported ferroelectric FET devices with 2D ferroelectric materials as the channel materials or non-ferroelectric 2D semiconductor/ferroelectric materials as the channel materials/gate insulator[7,41,42,46-48]. CIPS: $CuInP_2S_6$; PZT: $Pb[Zr_{0.2}Ti_{0.8}]O_3$.

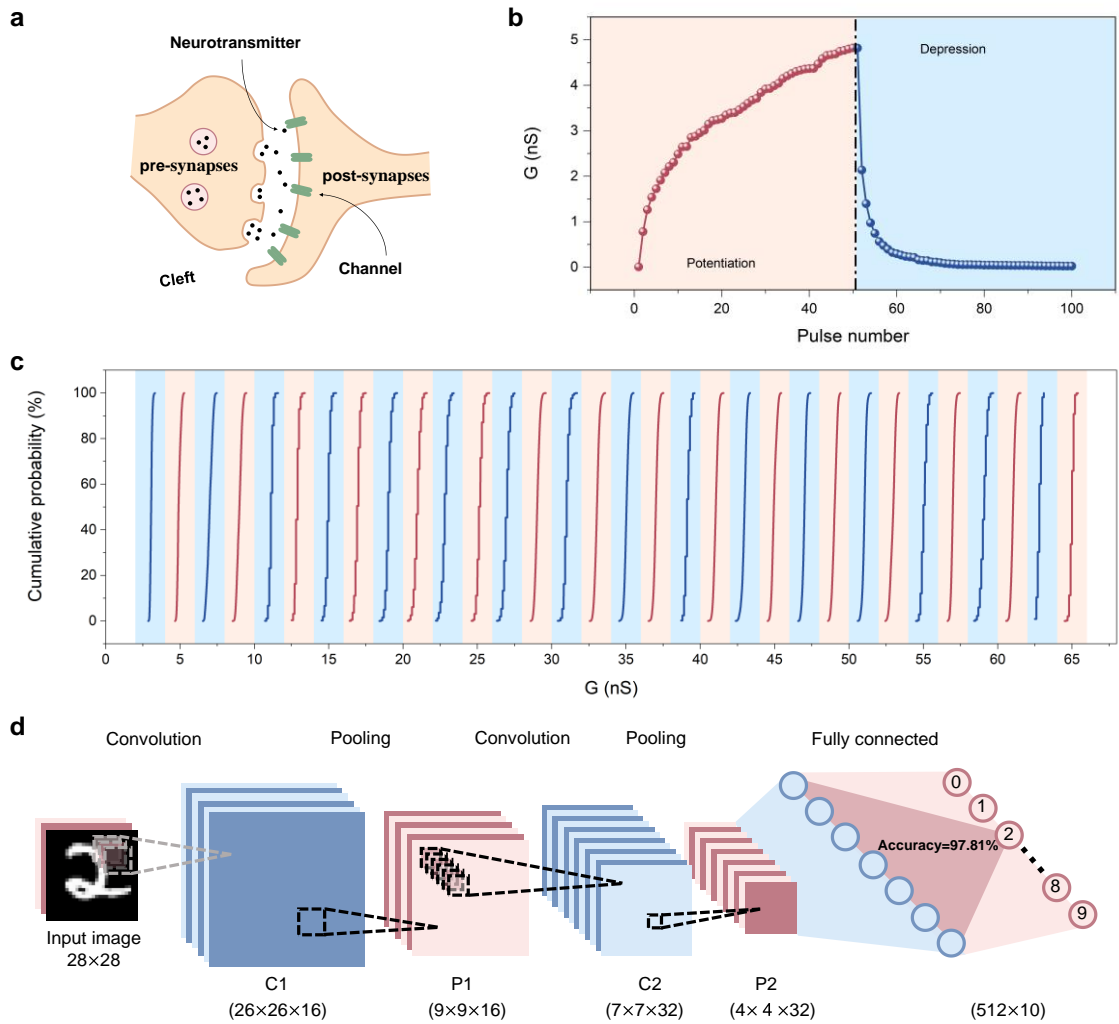

**Fig. 4 | Flexible sliding ferroelectric synapse. a**, Schematic of biological synapse. Signal transmission from the pre-synaptic terminal to the post-synaptic terminal is realized through ion channel by releasing neurotransmitters. **b**, The synaptic weight PSC of a flexible SFeM versus gate pulse numbers, indicating the long-term potentiation and depression characteristics. **c**, Cumulative probability of a flexible SFeM with respect to 32 independent drain-to-source conductance states. The conductance is programmed by applying different numbers of gate voltage pulses with 5 V amplitude and 10 ms width. **d**, Structure of the five-layer CNN used for MNIST image recognition containing two convolutional layers (C1 and C2), two maximum-pooling layers (P1 and P2), and a fully-connected layer. The input is a 28×28 grey scale (8bit) digit image. Bottom labels provide the feature map dimension (height × width × channel depth) or the vector dimension.

**Reference**

1    Dawber, M., Rabe, K. M. & Scott, J. F. Physics of thin-film ferroelectric oxides. *Rev. Mod. Phys.* **77**, 1083-1130 (2005).

2    Du, L. *et al.* Engineering symmetry breaking in 2D layered materials. *Nat. Rev. Phys.* **3**, 193-206 (2021).

3    Scott, J. F. Applications of Modern Ferroelectrics. *Science* **315**, 954-959 (2007).

4    Khan, A. I., Keshavarzi, A. & Datta, S. The future of ferroelectric field-effect transistor technology. *Nat. Electron.* **3**, 588-597 (2020).

5    Zhang, D., Schoenherr, P., Sharma, P. & Seidel, J. Ferroelectric order in van der Waals layered materials. *Nat. Rev. Mater.* **8**, 25-40 (2023).

6   Wang, C., You, L., Cobden, D. & Wang, J. Towards two-dimensional van der Waals ferroelectrics. *Nat. Mater.* **22**, 542-552 (2023).

7   Si, M. *et al.* A ferroelectric semiconductor field-effect transistor. *Nat. Electron.* **2**, 580-586 (2019).

8   Yang, S. Y. *et al.* Above-bandgap voltages from ferroelectric photovoltaic devices. *Nat. Nanotechnol.* **5**, 143-147 (2010).

9   Wu, J. *et al.* High tunnelling electroresistance in a ferroelectric van der Waals heterojunction via giant barrier height modulation. *Nat. Electron.* **3**, 466-472 (2020).

10  Si, M. *et al.* Steep-slope hysteresis-free negative capacitance $MoS_2$ transistors. *Nat. Nanotechnol.* **13**, 24-28 (2018).

11  Liu, F. *et al.* Room-temperature ferroelectricity in $CuInP_2S_6$ ultrathin flakes. *Nat. Commun.* **7**, 12357 (2016).

12  Fei, Z. *et al.* Ferroelectric switching of a two-dimensional metal. *Nature* **560**, 336-339 (2018).

13  Chang, K. *et al.* Discovery of robust in-plane ferroelectricity in atomic-thick SnTe. *Science* **353**, 274-278 (2016).

14  Zheng, Z. *et al.* Unconventional ferroelectricity in moiré heterostructures. *Nature* **588**, 71-76 (2020).

15  Li, L. & Wu, M. Binary Compound Bilayer and Multilayer with Vertical Polarizations: Two-Dimensional Ferroelectrics, Multiferroics, and Nanogenerators. *ACS Nano* **11**, 6382-6388 (2017).

16  Yasuda, K., Wang, X., Watanabe, K., Taniguchi, T. & Jarillo-Herrero, P. Stacking-engineered ferroelectricity in bilayer boron nitride. *Science* **372**, 1458-1462 (2021).

17  Vizner Stern, M. *et al.* Interfacial ferroelectricity by van der Waals sliding. *Science* **372**, 1462-1466 (2021).

18  Woods, C. R. *et al.* Charge-polarized interfacial superlattices in marginally twisted hexagonal boron nitride. *Nat. Commun.* **12**, 347 (2021).

19  Weston, A. *et al.* Interfacial ferroelectricity in marginally twisted 2D semiconductors. *Nat. Nanotechnol.* **17**, 390-395 (2022).

20  Wang, X. *et al.* Interfacial ferroelectricity in rhombohedral-stacked bilayer transition metal dichalcogenides. *Nat. Nanotechnol.* **17**, 367-371 (2022).

21  Rogée, L. *et al.* Ferroelectricity in untwisted heterobilayers of transition metal dichalcogenides. *Science* **376**, 973-978 (2022).

22  Meng, P. *et al.* Sliding induced multiple polarization states in two-dimensional ferroelectrics. *Nat. Commun.* **13**, 7696 (2022).

23  Deb, S. *et al.* Cumulative polarization in conductive interfacial ferroelectrics. *Nature* **612**, 465-469 (2022).

24  Sung, J. *et al.* Broken mirror symmetry in excitonic response of reconstructed domains in twisted $MoSe_2/MoSe_2$ bilayers. *Nat. Nanotechnol.* **15**, 750-754 (2020).

25  Wu, F. *et al.* Coupled Ferroelectricity and Correlated States in a Twisted Quadrilayer $MoS_2$ Moiré Superlattice. *Chin. Phys. Lett.* **40**, 047303 (2023).

26  Du, L. *et al.* Moiré photonics and optoelectronics. *Science* **379**, eadg0014 (2023).

27  Yang, L., Ding, S., Gao, J. & Wu, M. Atypical Sliding and Moiré Ferroelectricity in Pure Multilayer Graphene. *Phys. Rev. Lett.* **131**, 096801 (2023).

28  Ko, K. *et al.* Operando electron microscopy investigation of polar domain dynamics in twisted van der Waals homobilayers. *Nat. Mater.* **22**, 992-998 (2023).

29  Li, N. *et al.* Large-scale flexible and transparent electronics based on monolayer molybdenum disulfide field-effect transistors. *Nat. Electron.* **3**, 711-717 (2020).

30  Tang, J. *et al.* Low power flexible monolayer $MoS_2$ integrated circuits. *Nat. Commun.* **14**, 3633 (2023).

31  Liu, C. *et al.* Two-dimensional materials for next-generation computing technologies. *Nat. Nanotechnol.* **15**, 545-557 (2020).

32  Zhu, K. *et al.* The development of integrated circuits based on two-dimensional materials. *Nat. Electron.* **4**, 775-785 (2021).

33  Das, S. *et al.* Transistors based on two-dimensional materials for future integrated circuits. *Nat. Electron.* **4**, 786-799 (2021).

34  Yang, D. *et al.* Non-volatile electrical polarization switching via domain wall release in 3R-$MoS_2$ bilayer. *arXiv:2311.12126* (2023).

35  Yang, D. *et al.* Spontaneous-polarization-induced photovoltaic effect in rhombohedrally stacked $MoS_2$. *Nat. Photon.* **16**, 469-474 (2022).

36  Qi, J. *et al.* Stacking-Controlled Growth of rBN Crystalline Films with High Nonlinear Optical Conversion Efficiency up to 1%. *Adv. Mater.* **n/a**, 2303122, doi:https://doi.org/10.1002/adma.202303122 (2023).

37  Wang, Q. *et al.* Layer-by-layer epitaxy of multi-layer $MoS_2$ wafers. *Natl Sci. Rev.* **9**, nwac077 (2022).

38  Du, L. *et al.* Giant Valley Coherence at Room Temperature in 3R $WS_2$ with Broken Inversion Symmetry. *Research* **2019**, 6494565 (2019).

39  Zhao, M. *et al.* Atomically phase-matched second-harmonic generation in a 2D crystal. *Light Sci. Appl.* **5**, e16131-e16131 (2016).

40  Yang, T. H. *et al.* Ferroelectric transistors based on shear-transformation-mediated rhombohedral-stacked molybdenum disulfide. *Nat. Electron.*, doi:10.1038/s41928-023-01073-0 (2023).

41  Wang, S. *et al.* Two-dimensional ferroelectric channel transistors integrating ultra-fast memory and neural computing. *Nat. Commun.* **12**, 53 (2021).

42  Wang, L. *et al.* Exploring Ferroelectric Switching in α-$In_2Se_3$ for Neuromorphic Computing. *Adv. Funct. Mater.* **30**, 2004609 (2020).

43  Li, W. *et al.* Approaching the quantum limit in two-dimensional semiconductor contacts. *Nature* **613**, 274-279 (2023).

44  Sebastian, A., Le Gallo, M., Khaddam-Aljameh, R. & Eleftheriou, E. Memory devices and applications for in-memory computing. *Nat. Nanotechnol.* **15**, 529-544 (2020).

45  Gao, L. *et al.* Intrinsically elastic polymer ferroelectric by precise slight cross-linking. *Science* **381**, 540-544 (2023).

46  Wang, X. *et al.* Van der Waals engineering of ferroelectric heterostructures for long-retention memory. *Nat. Commun.* **12**, 1109 (2021).

47  Kim, K.-H. *et al.* Scalable CMOS back-end-of-line-compatible AlScN/two-dimensional channel ferroelectric field-effect transistors. *Nat. Nanotechnol.* **18**, 1044-1050 (2023).

48  Ko, C. *et al.* Ferroelectrically Gated Atomically Thin Transition-Metal Dichalcogenides as Nonvolatile Memory. *Adv. Mater.* **28**, 2923-2930 (2016).

49  Shi, Y. *et al.* Electronic synapses made of layered two-dimensional materials. *Nat. Electron.* **1**, 458-465 (2018).

50  Kamaei, S. *et al.* Ferroelectric gating of two-dimensional semiconductors for the integration of steep-slope logic and neuromorphic devices. *Nat. Electron.* **6**, 658-668 (2023).

51  Yan, M. *et al.* Ferroelectric Synaptic Transistor Network for Associative Memory. *Adv. Electron. Mater.* **7**, 2001276 (2021).

52  Migliato Marega, G. *et al.* Logic-in-memory based on an atomically thin semiconductor. *Nature* **587**, 72-77 (2020).

53  Jung, S. *et al.* A crossbar array of magnetoresistive memory devices for in-memory computing. *Nature* **601**, 211-216 (2022).

54  Kresse, G. & Hafner, J. Ab initio molecular dynamics for liquid metals. *Phys. Rev. B* **47**, 558-561 (1993).

55  Kresse, G. & Hafner, J. Ab initio molecular-dynamics simulation of the liquid-metal--amorphous-semiconductor transition in germanium. *Phys. Rev. B* **49**, 14251-14269 (1994).

56  Perdew, J. P., Burke, K. & Ernzerhof, M. Generalized Gradient Approximation Made Simple. *Phys. Rev. Lett.* **77**, 3865-3868 (1996).

57  Blöchl, P. E. Projector augmented-wave method. *Phys. Rev. B* **50**, 17953-17979 (1994).

58  Monkhorst, H. J. & Pack, J. D. Special points for Brillouin-zone integrations. *Phys. Rev. B* **13**, 5188-5192 (1976).

59  Wang, V., Xu, N., Liu, J.-C., Tang, G. & Geng, W.-T. VASPKIT: A user-friendly interface facilitating high-throughput computing and analysis using VASP code. *Comput. Phys. Commun.* **267**, 108033 (2021).